\documentstyle[12pt,epsf]{article}
\newcommand{\plot}[1]
{\begin{center}
\epsfysize=12cm 
\vspace{-5mm}
\parbox{\epsfxsize}{\epsffile{#1}}
\vspace{5mm}
\end{center}}
\def\tm{$^{\mbox{\sc tm}}$}

\begin{document} 
\begin{titlepage}
\title{ALARMING OXYGEN DEPLETION CAUSED BY HYDROGEN COMBUSTION 
AND FUEL CELLS AND THEIR RESOLUTION BY MAGNEGAS$^{TM}$}
\author{\bf Ruggero Maria Santilli\\[1cm]
R\&D Director, USMagnegas, Inc.\\
13100 Belcher Road, Largo, FL 33773, U.S.A.\\
\normalsize 
Tel.: +1-727-507 9520, Fax: +1-727-507 8261, E-mail: \mbox{ibr@gte.net}}
\date{\normalsize Contributed paper,\\
International Hydrogen Energy Forum 2000,\\ 
Munich, Germany, September 11-15, 2000}
\maketitle

\abstract{
We recall that hydrogen combustion does resolve the environmental
problems of fossil fuels due to excessive emission of carcinogenic
substances and carbon dioxide. However, hydrogen combustion implies the
permanent removal from our atmosphere of directly usable oxygen, a
serious environmental problem called oxygen depletion, since the
combustion turns oxygen into water whose separation to restore the
original oxygen is prohibitive due to cost. We then show that a
conceivable global use of hydrogen in complete replacement of fossil
fuels would imply the permanent removal from our atmosphere of
2.8875$\times$10$^7$ metric tons O$_2$/day. 
Fuel cells are briefly discussed to point
out similarly serious environmental problems, again, for large uses. We
propose the possibility of resolving these problems by upgrading
hydrogen to the new combustible fuel called magnegas\tm{}, whose chemical
structure is composed by the new chemical species of magnecules, whose
energy content and other features are beyond the descriptive capacities
of quantum chemistry. In fact, magnegas contains up to 50\% hydrogen,
while having combustion exhaust with: 1) a positive oxygen balance
(releasing more oxygen in the exhaust than that used in the combustion);
2) no appreciable carcinogenic or toxic substances;  3) considerably
reduced carbon dioxide as compared to fossil fuels; 4) considerably
reduced nitrogen oxides; and 5) general reduction of pollutants in the
exhaust up to 96\% of current EPA standards. We also discuss the
possibility of further reducing carbon dioxide via suitable disposable
sponges in the exhaust system, as well as the further reduction of
nitrogen oxides with more efficient engine cooling and other means. The
analysis therefore indicates that magnegas combustion exhaust already is
dramatically below EPA standards, while the achievement of a completely
clean exhaust is within technological reach. Therefore, magnegas appears
to be an excellent upgrading of hydrogen, both, for direct combustion
and for use in fuel cells. We finally indicate that one of the best
applications of the new technology is that of processing crude oil in
the magnegas reactors, by yielding a fuel dramatically cleaner than
gasoline, at a cost smaller than that via refineries. In conclusion,
crude oil, hydrogen and fuel cells remain indeed fully admissible in
this new era of environmental concern, provided that they are treated
via a basically new technology whose quantitative study requires a new
chemistry, called hadronic chemistry [1-5].}
\end{titlepage}

\setcounter{page}{2}
 As is well known, gasoline combustion requires atmospheric oxygen which
is then turned into CO$_2$ and various HydroCarbon (HC). In turn, CO$_2$ is 
recycled by plants via the known reaction, 
H$_2$O + CO$_2$ + ($h\nu$) $\to$ O$_2$ + \mbox{(-(CH$_2$O)-)}, 
which restores oxygen in the atmosphere. Essentially this
was the scenario at the beginning of the 20-th century. The same
scenario at the beginning of the 21-st century is dramatically
different, because forests have rapidly diminished while we have reached
the following unreassuring  daily consumption of crude oil:\\

\noindent
74.18 million of barrel per day\\
= (74.18 million barrels/24h)$\times$(55 gallons/barrel)\\
= 4.08$\times$10$^9$ gallons/24h  \\
= 1.54$\times$10$^{13}$ cc/24h (using 4 quarts/gallon and 946 cc/quart)
                                                            \hfill (1) \\
= (4.08$\times$10$^9$ gallons)$\times$(4 qrt./gallon)x(946 cc/qrt.)/day \\
= 1.5438$\times$10$^{13}$ cc/day\\
= (1.5438$\times$10$^{13}$ cc/day)$\times$(0.7028 grams/cc)\\
= 1.0850$\times$10$^{13}$ grams octane/day\\
= (1.0850$\times$10$^{13}$ grams)/(114.23 grams/mole)\\
= 9.4984$\times$10$^{10}$ moles n-octane/day,\\

\noindent
(see, e.g., http://www.eia.doe.gov/emeu/international/energy.html)
where we have replaced, for simplicity, crude oil with a straight chain
of n-octanes CH$_3$-(CH$_2$)$_6$-CH$_3$ with the known density of 0.7028 g/cc 
at 20$^o$ C. It should be indicated that data (1) do not include the additional
large use of natural gas and coals, which would bring the daily
combustion of all fossil fuel to the equivalent of about 120 million
barrels of crude oil per day.

 The primary environmental problems caused by the above disproportionate
consumption of fossil fuel per day are the following:

 1) Excessive emission of carcinogenic and other toxic substances in the
combustion exhaust. It is well known by experts that gasoline combustion
releases in our atmosphere the largest percentage of carcinogenic and
other toxic substances as compared to any other source. The terms
"atmospheric pollution" are an euphemism for very toxic breathing.

 2) Excessive release of carbon dioxide. It is evident that, under the
very large daily combustion (1), plants cannot recycle the entire
production of CO$_2$, thus resulting in an alarming increase of CO$_2$ in our
atmosphere, an occurrence known as green house effect. In fact,  by
using the known reaction 
C$_8$H$_{18}$ + (25/2)O$_2$ $\to$ 8CO$_2$ + 9H$_2$O, 
we have the
following alarming daily production of CO$_2$ from fossil fuel combustion:\\

\noindent
(9.4984$\times$10$^{10}$ moles C$_8$H$_{18}$)$\times$(8/1)/day \\
= 7.5987$\times$10$^{11}$ moles CO$_2$/day \\
= (7.5987$\times$10$^{11}$ moles)$\times$(0.044 kg/mole)/day\hfill (2)\\
= 3.3434$\times$10$^{10}$ kg/day \\
= (3.3434$\times$10$^{10}$ kg/day)/(1000 kg/metric ton) \\
= 3.3434$\times$10$^7$ metric tons/day.\\

\noindent
 It is evident that plants cannot possibly recycle such a
disproportionate amount of daily production of CO$_2$. This has implied a
considerable increase of CO$_2$ in our atmosphere which can be measured by
any person seriously interested in the environment via the mere purchase
of a CO$_2$ meter, and then compare current readings of CO$_2$ with standard
values on record, e.g., the percentage of CO$_2$ in our atmosphere at sea
level in 1950 was 0.033\% $\pm$ 0.01\% (see, e.g., 
{\it Encyclopedia Britannica}
of that period). Along these lines, in our laboratory in Florida we
measure a thirty fold increase of CO$_2$ in our atmosphere over the
indicated standard. We assume the reader is aware of recent TV reports
of small areas in the North Pole containing liquid water, an occurrence
which has never been observed before. Increasingly catastrophic
climactic events are known to everybody.

 3) Excessive removal of directly usable oxygen from our atmosphere, an
environmental problem of fossil fuel combustion, which is lesser known
than the green house effect, even among environmentalists, but
potentially more serious. The problem is called oxygen depletion, and
refers to the difference between the oxygen needed for the combustion
less that expelled in the exhaust. By using again the reaction 
C$_8$H$_{18}$ +(25/2)O$_2$ $\to$ 8CO$_2$ + 9H$_2$O 
and data (2), it is easy to obtain the
following additionally alarming daily use of oxygen for the combustion
of fossil fuel,\\

\noindent
(9.4984$\times$10$^{10}$ moles octane/day)$\times$
(12.5 moles O$_2$/1 mole octane) \\
= 1.1873$\times$10$^{12}$ moles of O$_2$/day \\
= (1.1873$\times$10$^{12}$ moles of O$_2$)$\times$(0.032 kg/mole O$_2$)
                                                           \hfill (3)\\
= 3.7994$\times$10$^{10}$ kg O$_2$/day \\
= 3.7994$\times$10$^7$ metric tons/day.\\

\noindent
 Again, this large volume of oxygen is turned by the combustion into CO$_2$
of which only an unknown part is recycled by plants into usable oxygen.
Thus, the actual and permanent oxygen depletion caused by fossil fuel
combustion in our planet is currently unknown. However, it should be
indicated that the very existence of the green house effect is
unquestionable evidence of oxygen depletion, because we are dealing
precisely with the quantity of CO$_2$ which has not been re-converted into
O$_2$ by plants.

 Oxygen depletion is today measurable by any person seriously interested
in the environment via the mere purchase of an oxygen meter, measure the
local percentage of oxygen, and then compare the result to standards on
record, e.g., the oxygen percentage in our atmosphere at sea level in
1950 was 20.946\% $\pm$ 0.02\% (see, e.g., {\it Encyclopedia Britannica} 
of that period). 
Along these lines, in our laboratory in Florida we measure a local oxygen 
depletion of 3\%-5\%. Evidently, bigger oxygen depletions are
expected for densely populated areas, such as Manhattan, London, and
Tokyo, or at high elevation. We assume the reader is aware of the recent
decision by U.S. airlines to lower the altitude of their flights despite
the evident increase of cost. This decision has been apparently
motivated by oxygen depletion, e.g., fainting spells due to insufficient
oxygen suffered by passengers during flights at previous higher
altitudes.

 The purpose of this note is to indicate that, whether used for direct
combustion or in fuel cells, hydrogen combustion does not release
carcinogenic and carbon dioxide in the exhaust, but causes an alarming
oxygen depletion which is considerably bigger than that caused by fossil
fuel combustion for the same energy. This depletion is due to to the
fact that gasoline combustion turns oxygen into CO$_2$ part of which is
recycled by plants into O$_2$, while hydrogen combustion turns atmospheric
oxygen into H$_2$O. This process permanently removes oxygen from our planet
in a directly usable form due to the excessive cost of water separation
to restore the original oxygen balance.
 By assuming, for simplicity, that gasoline is solely composed of one
octane  C$_8$H$_{18}$, thus ignoring other isomers, the combustion of 
one mole
of H$_2$ gives 68.32 Kcal, while the combustion of one mole of octane
produces 1,302.7 Kcal. Thus, we need 19.07 = 1302.7/68.32 moles of H$_2$
to produce the same energy of one mole of octane.

 In turn, the combustion of 19.07 moles of H$_2$ requires 9.535 moles of
O$_2$, while the combustion of one mole of octane requires 12.5 moles of
O$_2$. Therefore, on grounds of the same energy release, the combustion of
hydrogen requires less oxygen than gasoline (about 76\% of the oxygen
consumed by the octane).

 The alarming oxygen depletion occurs, again, because of the fact that
the combustion of hydrogen turns oxygen into water, by therefore
permanently removing usable oxygen from our planet. When used in modest
amounts, the combustion of hydrogen constitutes no appreciable
environmental problem. However, when used in large amounts, the
combustion of hydrogen is potentially catastrophic on environmental
grounds, because oxygen is the foundation of life.

 At the limit, a global use of hydrogen as fuel in complete replacement
of fossil fuels would render our planet uninhabitable in a short period
of time. In fact, such a vast use of hydrogen would imply the permanent
removal from our atmosphere of 76\% of the oxygen currently consumed to
burn fossil fuels, i.e., from Eqs. (2) and (3), we have the following
permanent oxygen depletion due to global hydrogen combustion:\\

\indent $\qquad\qquad$
76\% oxygen used for fossil fuel combustion \hfill (4)\\
\indent $\qquad\qquad$
= 2.8875$\times$10$^7$ metric tons O$_2$ depleted/day,\\

\noindent
which would imply the termination of any life on Earth within a few
months.

 Predictably, the above feature of hydrogen combustion has alarmed
environmental groups, labor unions, and other concerned people. As an
illustration, calculations show that, in the event all fuels in
Manhattan were replaced by hydrogen, the local oxygen depletion would
cause heart failures, with evident large financial liabilities and legal
implications for hydrogen suppliers.
 An inspection of fuel cells reveals essentially the same scenario. If
hydrogen is used as fuel we have the above indicated oxygen depletion.
If, instead, we use fossil fuels in fuel cells, we are back to
essentially the original problems caused by fossil fuel combustions.
 The main open issue created by the above scenario is: since pure
hydrogen is potentially catastrophic on a large scale use whether as
direct fuel or in fuel cells, how can hydrogen be upgraded to a form
avoiding the oxygen depletion? It is easy to see that this question does
not admit an industrially and environmentally acceptable answer via the
use of conventional gases. For instance, the addition of CO to H$_2$ in a
50-50 mixture would leave the oxygen depletion unchanged. In fact, each
of the two reactions, 
H$_2$ + (1/2) O$_2$ $\to$ H$_2$ and 
CO + (1/2) O$_2$ $\to$ CO$_2$,
requires 1/2 mole of O$_2$.Therefore, the 50-50 mixture of H$_2$ and CO would
also require 1/2 mole of O$_2$, exactly as it is the case for the pure H$_2$.

 After studying the above problems for years, the only answer known to
this author is that of upgrading hydrogen into a new combustible gas,
called magnegas\tm{} [1] (international patents pending), which is
produced as a by-product  in the recycling of liquid waste (such as
automotive antifreeze and oil waste, city and farm sewage, etc.) or the
processing of carbon-rich liquids (such as crude oil, etc.). The new
technology, called PlasmaArcFlow\tm{} (international patents pending), is
essentially based on flowing liquids through a submerged electric arc
with at least one carbon electrode. The arc essentially decomposes the
liquid molecules into a plasma at 7,000$^o$~F composed of mostly ionized H,
O and C atoms, plus solid precipitates. The technology then controls the
recombination of H, O and C into a combustible gas with a new chemical
species, tentatively called magnecules\tm{} [2], which is currently under
study.

 A first peculiarity of magnegas\tm{} nonexistent in other gases, is that,
following numerous tests in analytic laboratories, its chemical
structure cannot be identified via conventional Gas Chromatographic Mass
Spectrometric (GC-MS) measurements, since it results to be constituted
by large clusters (all the way to 1,000 a.m.u. in molecular weight)
which remain completely unidentified by the MS. The chemical structure
of magnegas is equally unidentifyable via InfraRed Detectors (IRD),
because the new clusters composing magnegas\tm{} have no IR signature at
all, thus suggesting a bond of non-valence type (because these large
clusters cannot possibly be all symmetric).
 Moreover, the IR signature of conventional molecules such as CO and CO$_2$
result to be mutated with the appearance of new peaks, which evidently
indicate new internal bonds. These features establish that magnegas has
an energy content considerably bigger than that predicted by quantum
chemistry, since it can store energy in three different levels:
magnecules, molecules, and new internal molecular bonds. As a result,
the combustion of conventional fuels can be conceived as a singlet
rocket firing, while the combustion of ,magnegas can be referred to the
burning of a multi-stage rocket, with intriguing new features.
 In vies of the above occurrences, quantitative scientific studies of
magnegas are, therefore, beyond the capabilities of quantum chemistry. A
broader theory suitable for scientific studies of the new chemical
species and the combustion of the new gas has been developed by 
R.~M.~Santilli and D.~D.~Shillady under the name of hadronic chemistry [3, 4]
(see also papers [5]).

 Scans of the same sample of magnegas at different times shows different
magnecules, a phenomenon called magnecule mutation. The effect is
expected to be due to collisions among magnecules, resulting
fragmentations due to their large size, and their subsequent
recombinations with other fragments. This results in macroscopic changes
of the MS peaks for the same gas under the same GC-MS test, only
conducted at different times. These mutations have identified the
presence in the clusters of individual atoms of H, O and C, plus
ordinary molecules H$_2$, CO, and O$_2$ [2, 3]. The estimated conventional
composition of magnegas consists of about 40\%-45\% hydrogen, 55\%-60\%
carbon monoxide, the rest being composed by traces of oxygen and carbon
dioxide. Evidently, small traces of light HC are possible in ppm, but
no heavy HC is possible  in magnegas since the gas is created at 
7,000$^o$~F 
of the electric arc, as confirmed by the lack of activation of
catalytic converters during the combustion.
 As a working hypothesis in the absence of a more accurate knowledge, it
is conjectured that the very intense magnetic fields in the microscopic
vicinity of 1,000-3,000 DC Amps of the submerged electric arc (which can
be as high at 10$^{14}$ Gauss at distances of 10$^{-8}$ cm) cause 
a polarization
of the orbits of at least the valence electrons from a spherical into a
toroidal configuration, resulting in strong magnetic fields estimated to
be 1,415 times nuclear magnetic fields [2, 5a]. It is then expected that
strongly polarized individual atoms and molecules bond together like
little magnets, resulting in clusters which are stable at ordinary
conditions. Since the new bonds do not appear to be of valence type (or
any of its variations), they can only be of electric, magnetic, or
electromagnetic nature. The new clusters are called magnecules because
of the dominance of magnetic over other effects in their creation, while
electric effects are generally unstable, and often repulsive (as it is
the case of ions).

 Besides direct calculations [2, 5a], the  magnetic polarization of the
atoms and molecules constituting magnegas is further supported by a
number of indirect effects, such as the capability of magnegas\tm{} to
stick to instruments walls, called magnecule adhesion. As an
illustration, following the removal of magnegas from a GC-MS and its
conventional flushing, the background preserves all the anomalous peaks
of magnegas. This occurrence can only be interpreted numerically via
adhesion due to induced magnetic polarization, and not via
electrostatic, coordination, and other effects.
 Mutatis mutandae, stable clusters can only exist under a sufficiently
strong attractive force, which must be numerically identified for a
model to have sufficient depth. Among all possible non-valence bonds,
the magnetic attraction among polarized valence orbits is the only model
available at this writing with a concrete attractive bond,  while all
other models lack such an identification (as it is the case for electric
effects, coordination effects, co-valence, etc.). Due to the
implications here at stake, the study of alternative structure of the
new clusters in magnegas is warmly recommended, provided that, again,
the attractive force creating the clusters is specifically and
numerically identified, and models based on pure nomenclatures are
avoided.

 Even though the chemical structure of magnegas escapes current quantum
chemical knowledge, its combustion exhaust has a conventional chemical
structure, because the exhaust temperature is beyond the Curie point of
magnecules. As a result, all magnecules and other anomalies are removed
by the combustion.  Following numerous tests, including various
conversions of automobiles to run on magnegas, we have the following
combustion exhaust of magnegas measured before the catalytic converter,
in percentages:\\

\begin{center}
\begin{tabular}{ll}
\cline{1-2}\\
{\bf Water vapor}:     & 65\%-70\%\\
{\bf Oxygen}:          & 9.5\%-10.5\%\\
{\bf Carbon dioxide}:  & 6\%-8\%\\
{\bf Carbon monoxide}: & 0.00\%-0.01\%\\ 
{\bf Hydrocarbons}:    & minus 2 to minus 5 ppm\\
Rest atmospheric       &\\
\cline{1-2}
\end{tabular}
\hfill (5)\\
\end{center}
\medskip

\noindent
 As one can see, the upgrading of hydrogen into magnegas: 1) turns the
oxygen depletion caused by hydrogen combustion into a positive oxygen
balance (more oxygen in the exhaust than that used for the combustion)
2) emits no carcinogenic or toxic substance in the exhaust; and 3)
implies a significant reduction of carbon dioxide emission over fossil
fuels. In particular, magnegas exhaust meets the most stringent
governmental requirements without a catalytic converter while having a
positive oxygen balance.
 Preliminary magnegas exhaust measurements have been recently conducted
at the EPA Certified, Vehicle Certification Laboratory Liphardt \&
Associates of Long Island, New York, via the Varied Test Procedure (VTP)
as per Regulation 40-CFR, Part 86 on a Honda Civic Natural Gas Vehicle
VIN number 1HGEN1649WL000160, produced in 1998 (and purchased new in
1999) to operate with Compressed Natural Gas (CNG). This car was
converted by USMagnegas, Inc., Largo, Florida, to operate on Compressed
MagneGas (CMG) via: 1) the replacement of CNG with CMG; 2) the disabling
of the oxygen sensor (because magnegas has 20 times more oxygen in the
exhaust than natural gas); and 3) installing a multiple spark system (to
improve combustion); while leaving the rest of the car unchanged,
including its computer.

 The tests consisted of the conventional EPA routine for Regulation
40-CFR, Part 89, consisting of three separate and sequential tests
conducted on a computerized dynamometer, the first and the third tests
using the car at its maximal possible capability to simulate an up-hill
travel at 60 mph,  while the second test consists in simulating normal
city driving of the car. Three corresponding bags with the exhaust
residues are collected, jointly with a fourth bag containing atmospheric
contaminants. The final measurements expressed in grams/mile are given
by the average of the measurements on the three EPA test bags, less the
measurements of atmospheric pollutants in the fourth bag.
 The results of the above preliminary tests on magnegas exhaust are:\\

\begin{center}
\begin{tabular}{ll}
\cline{1-2}\\
{\bf Hydrocarbons}: &
0.026 gram/mile = 93.6\% reduction\\ &of the EPA standard of 0.41 gram/mile\\
{\bf Carbon monoxide}: &
0.262 grams/mile = 92.6\% reduction\\&of the EPA standard of 3.40 grams/mile\\
{\bf Nitrogen oxides}: &
0.281 gram/mile = 29.7\% reduction\\&of the EPA standard of 0.4 gm/mile\\
{\bf Carbon dioxide}: &
235 grams/mile - there is no EPA standard\\&on CO$_2$ at this moment\\
{\bf Oxygen}: &
not measured because not requested\\&in Regulation 40-CFR, Part 86\\
\cline{1-2}
\end{tabular}
\hfill (6)
\end{center}

\medskip
 The following comments are important for an appraisal of the above
results:

 1) Magnegas does not contain heavy HC since it is created at 7,000$^o$~F.
Therefore, the measured HC is expected to be due, at least in part, to
combustion of oil, either originating from magnegas compression pumps
(thus contaminating the gas), or from engine oil. New tests are under
way in which magnegas is filtered after compression, and all oils of
fossil fuels origin are replaced with synthetic oils.

 2) Carbon monoxide is fuel for magnegas (while being a combustion
product for gasoline). Therefore, any presence of CO in the exhaust is
evidence of insufficient combustion.

 3) The great majority of measurements (6) originate from the first and
third parts of the test at extreme performance, because, during ordinary
city traffic, magnegas exhaust is essentially pollutant free, as shown
in Figure~1.

 4) Nitrogen oxides are not due, in general, to the fuel (whether
magnegas or other fuel), but to the temperature of the engine, thus
being an indication of the quality of its cooling system. Therefore, for
each given fuel, including magnegas, NOx's can be decreased by improving
the cooling system and other means.

 5) Measurements (6) do not refer to the best possible combustion of
magnegas, but only to the combustion of magnegas in a vehicle whose
carburetion was developed for natural gas. Alternatively, the test was
primarily intended to prove the interchangeability of magnegas with
natural gas without any major automotive changes, while keeping
essentially the same performance and consumption. The measurements under
combustion specifically conceived for magnegas are under way, and will
be released in the near future. The main difference in the latter tests
is a considerable reduction in the emission of carbon dioxide for
certain technical reasons related to the magnegas combustion.

\begin{figure}
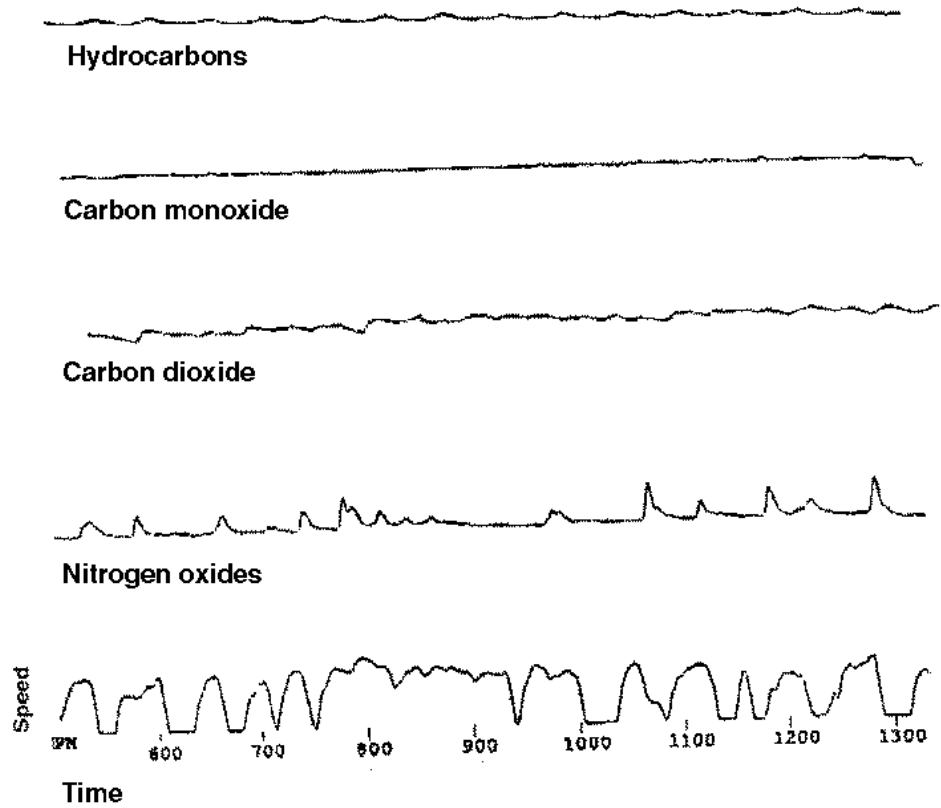

\plot{munich1.eps}
\caption{An illustration of the city part of the EPA test according to
Regulation 40-CFR, Part 86, conducted at the Vehicle Certification
Laboratory Liphardt \& Associates of Long Island, New York on a Honda
Civic Natural Gas Vehicle converted to magnegas. }
\label{Fig1}
\end{figure}

In Figure~1, the first three
diagrams illustrate the very low combustion emission of magnegas in city
driving, by keeping in mind that most of measurements (6) are due to the
heavy duty, hill climbing part of the EPA test. Even though 29.7\% of EPA
standard, the fourth diagram on nitrogen oxides is an indication of
insufficient cooling of the engine. The bottom diagram indicates the
simulated speed of the car versus time, where flat tracts simulate idle
portions at traffic lights. By keeping in mind: 1) the lack of (heavy)
hydrocarbon in magnegas (because produced at 7,000$^o$~F of the electric
arc); 2) the expectation of no appreciable carbon dioxide in the
magnegas exhaust under proper combustion (because CO is fuel for
magnegas); 3) the possible further reduction of carbon dioxide via
disposable sponges placed in the exhaust systems; 4) the decrease of
nitrogen oxides with a more efficient engine cooling and other
improvements; and 5) the positive oxygen balance of magnegas (not
measured in the test because not included in current EPA regulations);
the measurements depicted in this diagram indicate that the achievement
of a truly clean fuel is indeed within technological reach.

 We should also indicate considerable research efforts under way to
further reduce the CO$_2$ content via suitable cartridges of disposable
chemical sponges placed in the exhaust system. Admittedly, these
catalytic means generally implies the creation of acids harmful to the
human skin, if released in the environment. However, the ongoing
research aims at the chemical and/or technological resolution of these
problems. Additional research is under way via liquefied magnegas
obtained via catalytic or conventional liquefaction, which is expected
to have an anomalous energy content with respect to other liquid fuels,
and an expected, consequential decrease of pollutants. As a result of
these efforts, the achievement of an exhaust essentially free of CO$_2$
appears to be within technological reach.

 As a comparison for measurements (6), a similar (but different) Honda
car running on indolene (a version of gasoline) without affecting
performance was tested in the same laboratory with the same EPA
procedure, resulting in the following data:

\begin{center}
\begin{tabular}{ll}
\cline{1-2}\\
{\bf Hydrocarbons}:&
0.234 gram/mile  =  900\% of magnegas emission\\
{\bf Carbon monoxide}:&
1.965 gram/mile  =  750\% of magnegas emission\\
{\bf Nitrogen oxides}:&
0.247 gram/mile = 86\% of magnegas emission\hfill (7)\\
{\bf Carbon dioxide}:&
458.655 grams/mile = 195\% of magnegas emission\\
\cline{1-2}
\end{tabular}
\end{center}

\medskip
\noindent
which illustrates the environmental superiority of magnegas over
gasoline.
 The improvement of emission by magnegas over the above data are
evident.

 Other features favoring the upgrading of pure hydrogen into magnegas
are (international patents pending):

 1) magnegas is cost competitiveness with respect to fossil fuels (since
it is produced as a byproduct of an income-producing recycling);

 2) magnegas increases the energy content from about 300 BTU/cf for
hydrogen to about 800-900 BTU/cf (due to the new means of energy
storage);

 3) magnegas is more readily availability anywhere desired (since easily
transportable PlasmaArcFlow reactors as big as a desk produce up to
1,500 cf of magnegas per hour, i.e, a production in one hour sufficient
for about three hours city travel by a compact car);

 4) magnegas admits easier liquefaction, e.g., via Fischer-Tropsch
catalytic synthesis or conventional liquefaction (due to attractions
between magnecules);

 5) magnegas has a better penetration through membranes (due to measured
decreases of average molecular sizes of magnetically polarized
conventional molecules);

 6) magnegas can be used for any conventional fuel application,
including metal cutting, cooking, automotive use, etc.

 7) Magnegas can be used in fuel cells, by preserving its environmental
advantages.

 Above all, the magnegas\tm{} technology appears to permit an ultimate
merger of crude oil and hydrogen technologies. One of the best liquids
usable in the PlasmaArcFlow\tm{} reactors is crude oil, which is then
turned into a fuel much cleaner than gasoline (plus usable heat and
solid precipitates) at a cost visibly smaller than that that via huge
refineries. The fuel produced by the above new processing of crude oil
is 40\%-45\% hydrogen.

 In conclusion, crude oil, hydrogen, and fuel cells remain indeed fully
admissible in this new era of environmental concern, provided that they
are treated via a basically new technology whose quantitative study
requires a new chemistry, hadronic chemistry [1-5].

\section*{Acknowledgments} 

The author would like to thank D.~D.~Shillady, 
Chemistry Department, Virginia Commonwealth University, U.S.A., and 
A.~K.~Aringazin, Department of Theoretical Physics, Karaganda State
University, Kazakstan. Particular thanks are also due to all member of
USMagnegas, Inc., for invaluable assistance without which this paper
could not have seen the light of the day.

\end{document}